# Driving positron beam acceleration with coherent transition radiation

Zhangli Xu[1,2], Longqing Yi[3], Baifei Shen[1,4✉], Jiancai Xu[1✉], Liangliang Ji[1,5✉], Tongjun Xu[1], Lingang Zhang[1], Shun Li[1] & Zhizhan Xu[1]

Positron acceleration in plasma wakefield faces significant challenges, as the positron beam must be pre-generated and precisely coupled into the wakefield and, most critically, suffers from defocusing issues. Here we propose a scheme that utilizes laser-driven electrons to produce, inject, and accelerate positrons in a single setup. The high-charge electron beam from wakefield acceleration creates copious electron–positron pairs via the Bethe–Heitler process, followed by enormous coherent transition radiation due to the electrons' exiting from the metallic foil. Simulation results show that the coherent transition radiation field reaches up to tens of GV m$^{-1}$, which captures and accelerates the positrons to cut-off energy of 1.5 GeV with energy peak of 500 MeV (energy spread ~ 24.3%). An external longitudinal magnetic field of 30 T is also applied to guide the electrons and positrons during the acceleration process. This proposed method offers a promising way to obtain GeV fast positron sources.

[1] State Key Laboratory of High Field Laser Physics and CAS Center for Excellence in Ultra-intense Laser Science, Shanghai Institute of Optics and Fine Mechanics (SIOM), Chinese Academy of Sciences (CAS), Shanghai 201800, China. [2] Center of Materials Science and Optoelectronics Engineering, University of Chinese Academy of Sciences, Beijing 100049, China. [3] Department of Physics, Chalmers University of Technology, 41296 Gothenburg, Sweden. [4] Department of Physics, Shanghai Normal University, Shanghai 200234, China. [5] Center for Excellence in Ultra-intense Laser Science, Chinese Academy of Sciences, Shanghai 201800, China. ✉email: bfshen@mail.shcnc.ac.cn; jcxu@siom.ac.cn; jill@siom.ac.cn

Since the concept of laser wakefield acceleration (LWFA) being first proposed[1], electron acceleration driven by femtosecond relativistic laser pulse has made remarkable progress[2–4]. Electron energy has reached 7.8 GeV with small energy spread[5]. Furthermore, electron beams with short pulse duration of a few femtoseconds[6], small angular divergence of less than 1.5 mrad[7] or beam charge up to 10 nC[8] have been obtained. As the antimatter counterpart of electrons, positrons are novel sources for material science[9], laboratory astrophysics[10], and most of all, essential for future electron–positron colliders—an ultimate goal of laser-plasma acceleration[11]. Unlike electrons which naturally exist, positrons are to be generated in laboratories, usually via the Bethe–Heitler (BH) process[12] and the Breit–Wheeler process[13]. The former relies on the collision between energetic electrons with the high-Z atomic fields. Copious positron production has been demonstrated based on either ultraintense laser directly irradiating metallic foils[14–16] or sending the LWFA electron beams onto high-Z targets[17–20].

However, acceleration of pre-generated positrons is not as straightforward as it is for electrons. A key issue is that the plasma wakefield valid for electron acceleration defocuses positively charged particles. Several methods have been proposed from simulations to mitigate the problem via wakefield driven by electrons[21], hollow electron beam[22], two electron beams[23], vortex laser pulses[24], and in plasma channel[25]. These methods generally require external injection of well-prepared positrons, a prominent challenge even for the more mature electron acceleration. In experiments, positrons are able to gain energy (tens of MeVs to several GeVs) from positron driven wakefiled[26–28], which relies on tens of GeV dense positron beams only accessible in several large conventional accelerator facilities. Recently, laser-driven positron acceleration has been observed, where positrons created via the BH process in the high-Z foil gain ~MeV energies from the sheath field at the target rear surface[15,29].

Here, we propose an all-optical approach by coupling the production, injection, and long-distance acceleration to generate positrons beyond GeV. We find that, when the dense electron bunch from LWFA emits from copper target into vacuum, the induced strong coherent transition radiation (CTR) is capable of trapping and accelerating positrons for a long distance. The CTR field provides an acceleration gradient up to tens of GV m$^{-1}$, which is between that of the conventional accelerators (~100 MV m$^{-1}$)[30,31] and the plasma wakefield accelerators (~100 GV m$^{-1}$). Our combined simulations (EPOCH-Geant4-EPOCH) demonstrate that a quasi-monoenergetic position beam (energy spread ~24% at peak energy ~500 MeV) with cut-off energy of 1.5 GeV. The beam is highly collimated (2.5 mrad angular divergence) and short pulsed (duration ~10 fs). About 9.4 pC beam charge is obtained by a $1.37 \times 10^{20}$ W cm$^{-2}$, 223 J driving laser pulse. The acceleration mechanism in our scenario addresses the key challenges in laser-positron acceleration and provides a practical approach to develop all-optical GeV-level positron sources.

## Results and discussion
**Overview of the scheme.** A sketch of energetic positron generation setup is shown in Fig. 1a. It is composed of three stages as follows: stage I is the generation of high-charge, high-energy electrons in LWFA; stage II is the positron generation in the copper target via the routine BH process; and stage III is the acceleration process for positrons. We simulate the electron acceleration, CTR, and positron acceleration via the particle-in-cell (PIC) code EPOCH[32]. The BH process is analyzed using the Monte-Carlo code Geant4[33]. Connection between each process is made by conserving the beam features.

**Simulation results of positrons generation and acceleration.** To maximize the pair production events and the transition radiation intensity, in stage I we focus on generating high-charge electron beam at moderate kinetic energies. This is verified by two-dimensional (2D) PIC simulation code EPOCH. A circular-polarized laser pulse (wavelength $\lambda = 0.8$ μm, duration $\tau = 45$ fs) with super-Gaussian spatial distribution and $\sin^2$ temporal profile interacts with gas plasma with a density downramp. Detailed parameters are shown in Methods. This setup guarantees robust and efficient injection of electrons in a single pass. After the laser pulse transverses the density transition region, several nC of electrons are loaded. Further acceleration generates a beam of a continuous energy spectrum with cutoff energy of 2 GeV, as shown by black solid line in Fig. 1b. The total charge of electrons (>5 MeV within a divergence angle of 26.2 mrad) reaches about 3 nC, estimated by multiplying the 2D results by the transverse size of the electron beam. The full width at half-maximum (FWHM) transverse size of the electron beam is 15.5 μm and the longitudinal length is 4.5 μm (FWHM). The proposed approach also works for linear-polarized laser pulses, as long as a spatially well-defined high-charge electron beam is created.

The high-charge electron beam from stage I impinges onto a 1 mm-thick copper target set 0.5 mm away from the gas target, triggering the generation of high-energy Gamma photons via bremsstrahlung[34] and further production of electron–positron pairs from the BH process. In this case, the plasma collective effect is insignificant when the electron beam propagates through the target[35] and this process can be accomplished by the Monte-Carlo simulation code Geant4[33] (see "Methods"), which also provides the energy and angular spectrum of both electrons and positrons after they exit the copper target. For the sampled $10^7$ electrons, we obtain $5.03 \times 10^4$ positrons (generation efficiency ~ 0.5%). They show a Maxwellian energy spectrum of featured temperature $T_p = 24.1$ MeV, as shown by the black solid line in Fig. 1c. The positron beam has a divergence angle of 49.4 mrad (FWHM). We choose the 1 mm copper target thickness to make sure that substantial positrons are generated without significant energy loss for the electrons. The electron beam still contains 93.4% of its initial energy after the copper target. The beam is diverged due to the scattering. Its size is elongated to 44.8 μm (FWHM) in transverse and 5.3 μm (FWHM) longitudinally. The positron beam is in similar size as compared to the electron beam (42.5 μm × 5.3 μm), containing 12.1 pC beam charge according to the number ratio. Both electron and positron beams, leaving the copper target, co-propagate forward and overlap in space. Necessarily, the annihilation of positrons and electrons is negligible during the co-propagating process, as the lifetime of a positron is expressed as $\tau = 1/\left(\pi r_{e0}^2 c n_{e^-}\right)$, where $r_{e0}$ is electron classical radius and $n_{e^-}$ is the electron density where positrons located. In our case, $\tau \sim 0.135$ ms, positrons are able to propagate several kilometers without apparent annihilation.

In stage III, the electron beam exiting into vacuum from the rear surface of copper target drives intense CTR to accelerate positrons. The acceleration process is also verified by 2D PIC simulation (see "Methods"). Our PIC simulation confirms that when the electron beam emits into vacuum from the rear side of copper target, it induces strong CTR. The field co-propagates with the compound beam in vacuum and accelerates the forward-going positrons efficiently. The positron energy spectrum at different times in stage III are shown in Fig. 1c. A mono-energetic peak appears at the early beginning. After 400 mm propagation distance (1334.3 ps), the peak is accelerated to about 500 MeV. Energy spread is well reserved at about 24%. The cutoff energy reaches 1.5 GeV. This is accompanied by the gradual energy depletion for electrons, as shown in Fig. 1b.

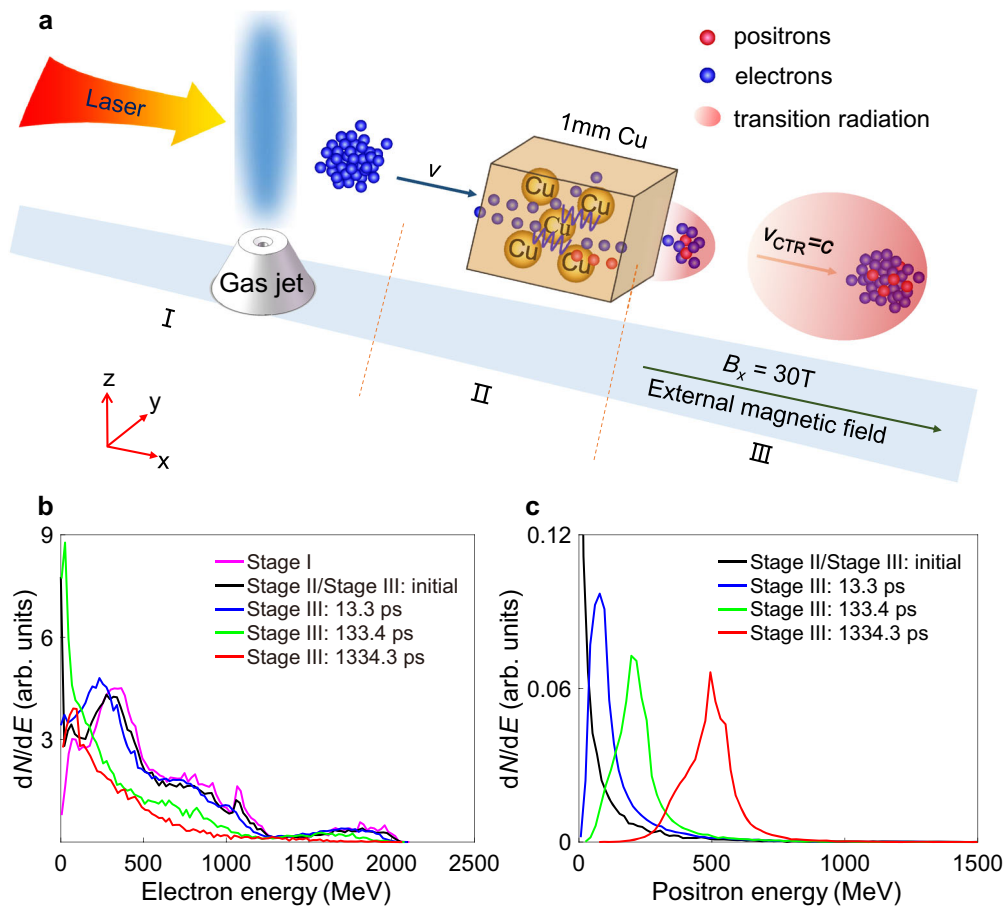

**Fig. 1 Proposed setup and simulation results of particles' energy spectra. a** Sketch of the proposed setup. A laser pulse is focused over the gas jet, which accelerates electrons (blue sphere) to ~500 MeV with beam charge up to ~nC. The electrons with velocity of $v$ further interact with the copper target. Positrons (red sphere) are generated via the Bethe–Heitler (BH) process and accelerated by longitudinal coherent transition radiation (CTR) field. The guiding magnetic field is 30 T. **b** Electron energy spectra after stage I, stage II/stage III (initial), and 13.3 ps (4 mm), 133.4 ps (40 mm), and 1334.3 ps (400 mm) in stage III. **c** The energy spectra of positrons after stage II/stage III (initial) and 13.3 ps (4 mm), 133.4 ps (40 mm), and 1334.3 ps (400 mm) in stage III.

Figure 2a plots the density evolution of positron beam in this acceleration process and indicates that the positron beam keeps its high density as ~$2 \times 10^{15}$ cm$^{-3}$ without significant increasing of the transverse size because of the focusing field. The beam is somehow compressed longitudinally. At 400 mm, the beam length of positrons decreases from 5.3 to 2.9 μm (FWHM), correspondingly to pulse duration of 9.7 fs. Some of the positrons, especially the ones located at the beam tail, undergo defocusing effect. The beam loses its charge by 2.7 pC in stage III. However, about 77% positrons are well preserved at the end of simulation. The focusing CTR field, as we will discuss later, results in good beam collimation. The angular divergence of the positron beam is 2.5 mrad (FWHM) and the normalized root-mean-squared beam emittance is 85.7 mm mrad. The phase-space distribution $p_x - x$ in Fig. 2b shows continuous acceleration for the confined positrons through the 400 mm propagation distance. Accordingly, the distributions of the longitudinal electric field $E_x$ at $x = 4$, 40, and 400 mm are given in Fig. 2c. The field shows a positive peak in the region occupied by electrons/positrons, followed by a decaying tail. One observes over 10 GV m$^{-1}$ acceleration field at $x = 4$ mm. It then declines gradually during propagation. Nonetheless, the field remains above GV m$^{-1}$ for the most part of the acceleration process.

One notices that the field $E_x$, shown by red solid line in Fig. 2b, decreases with $x$ when $x > x_0$ ($x_0$ is the place where $E_x$ peaks). For positrons mostly located at $x > x_0$, the less energetic particles fall behind but experience higher field and gain more acceleration than those with higher initial energies. Therefore, the energy chirp is eliminated, leading to a narrow energy spread width as the acceleration goes on. The effect also induces beam compression to form a compact bunch of high density and short duration. In our simulations, we have not reached the stage of phase reverse. The focusing field for positrons is estimated by $E_y - cB_z$, where $E_y$ is the transverse electric field and $B_z$ is the magnetic field in z direction, respectively. The 2D map and the lineout across the beam front are depicted in Fig. 2d and blue solid line in Fig. 2a, respectively. It has a peak value of 5 GV m$^{-1}$ at 4 mm and is radially inward in the region occupied by the positrons, a typical structure that provides efficient focusing for positrons. It is worth mentioning that the radiation is coherent[36], because the longitudinal length of electron beam (5.3 μm) is smaller than the radiation wavelength, which is ~9.6 μm obtained by transforming the field to the frequency regime.

**Theoretical calculation.** We start by studying the fields of an individual point charge in the case of the beam exiting a perfectly conducting plane at constant velocity. According to our simulation, a point charge $e$ exits an infinite perfectly conductor plane $x_0 = 20$ μm at constant velocity $v$. We shall also use spherical coordinates $(R, \theta, \varphi)$, with $R = (x^2 + y^2 + z^2)^{1/2}$, the polar angle measured from the $+x$ axis, and $\varphi$ the azimuth angle, as well as coordinates $(r, \varphi, x)$, with $r = (y^2 + z^2)^{1/2} = R\sin\theta$, and $x = R\cos\theta$. The fields components are obtained[37]

$$E_x = \frac{e}{4\pi\varepsilon_0\gamma^2}\left[\frac{x-x_0-vt}{S_-^3} - \frac{x-x_0+vt}{S_+^3}\right]u(ct-R) \\ - \frac{2\beta e \sin^2\theta}{4\pi\varepsilon_0 R(1-\beta^2\cos^2\theta)}\delta(ct-R), \quad (1)$$

$$E_r = \frac{e}{4\pi\varepsilon_0\gamma^2}\left[\frac{r}{S_-^3} - \frac{r}{S_+^3}\right]u(ct-R) + \frac{2\beta e \sin\theta\cos\theta}{4\pi\varepsilon_0 R(1-\beta^2\cos^2\theta)}\delta(ct-R), \quad (2)$$

where $S_- \equiv \sqrt{(vt-(x-x_0))^2 + r^2/\gamma^2}$, $S_+ \equiv \sqrt{(vt+(x-x_0))^2 + r^2/\gamma^2}$. The first terms of Eqs. (1) and (2), proportional to $u(ct-R)$, are the Lorentz-transformed Coulomb fields of the moving charge. They are confined to the spherical volume $R < vt$, and drop quickly to zero at $R = ct$. The second terms are the radiation fields, confined to the infinitely thin shell at $R = ct$. The radiation fields decay as $1/R$, whereas the Coulomb field decays as $1/R^2$. Then, the fields for a beam are constructed by summing those of elementary charges.

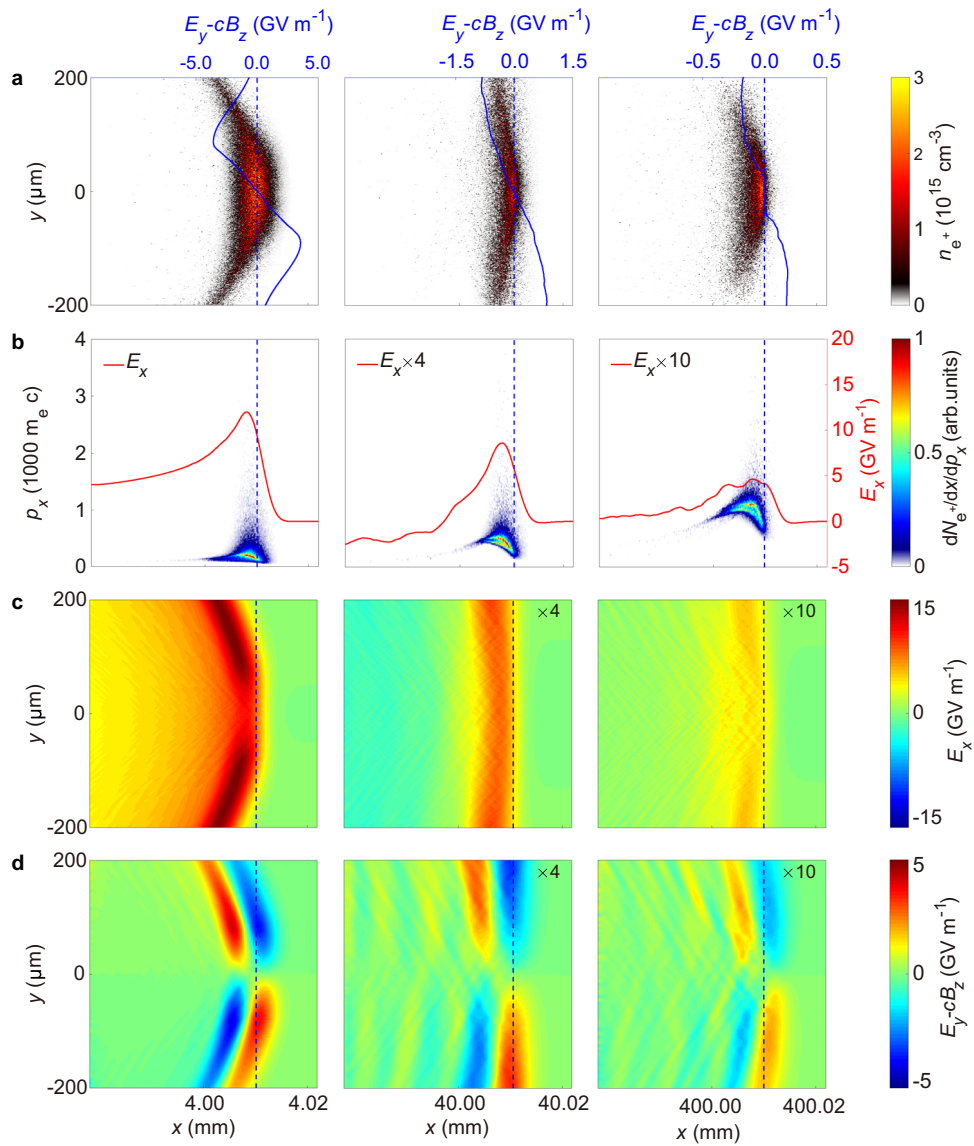

**Fig. 2 Simulation results of positrons' acceleration process. a** Positron beam density, **b** phase-space density of positrons, **c** longitudinal electric field $E_x$, and **d** transverse field at 4, 40, and 400 mm, respectively. The blue solid line in **a** represents transverse field at $x = 4.01, 40.01$, and 400.01 mm (denoted by the blue dashed line). The red solid line in **b** represents the on-axis longitudinal electric field. The longitudinal and transverse electric fields at 40 mm and 400 mm have been multiplied by four times and ten times, respectively. Simulation results at the beginning are shown in Supplementary Fig. 1 and Supplementary Note 1.

To verify the acceleration field caused by the CTR, which is three dimensional (3D) in Eqs. (1) and (2), we have done another 3D PIC simulation using the same beam parameters as in 2D simulation in stage III. Due to the limited computation recourses, we restrict the acceleration distance to be 2 mm. Simulations use a moving window with sizes of 40 μm × 200 μm × 200 μm divided into $200 \times 200 \times 200$ cells with 1 macro particle per cell for electrons and positrons. The simulation result of longitudinal electric field $E_x$ at 0.16 mm, plotted in Fig. 3a, shows that the maximum acceleration field reaches above 50 GV m$^{-1}$, The peak value and the field distribution matches the longitudinal field of the CTR shown in Fig. 3c, which is calculated by the second term of Eq. (1). We notice that the CTR field is two orders of magnitudes higher than that of Coulomb fields [shown in Fig. 3b, calculated by the first term of Eq. (1)]. As an estimation for the latter, we calculated the potential generated by electrons via

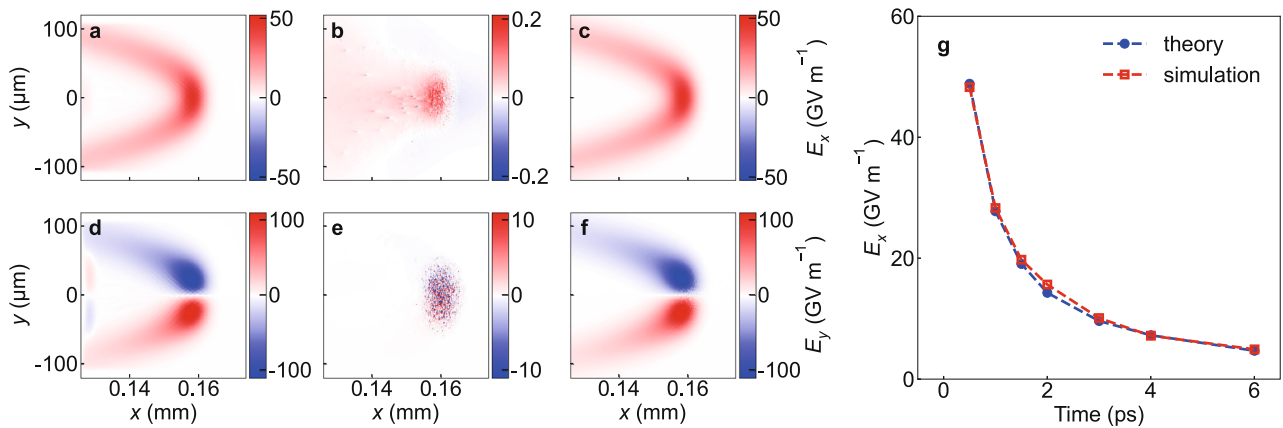

**Fig. 3 Comparison of simulation results with theory. a** Simulation results of longitudinal electric field at $x \sim 0.16$ mm. **b** Longitudinal Coulomb field calculated by first term of Eq. (1). **c** Longitudinal coherent transition radiation (CTR) field calculated by second term of Eq. (1). **d** Simulation results of transverse electric field at $x \sim 0.16$ mm. **e**, **f** The first and second terms of Eq. (2), representing the transverse Coulomb field and transverse CTR field, respectively. **g** Theoretical (blue solid circle) and simulation (red square) results of the maximum on-axis value of the longitudinal electric field at different times.

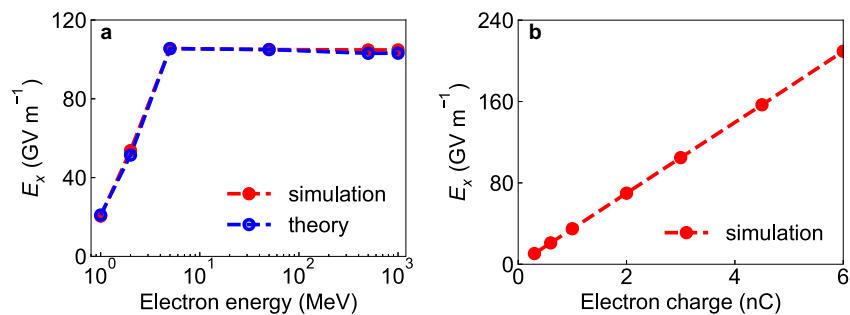

**Fig. 4 Scalings of longitudinal electric field versus the electron beam energy and charge. a** Simulation (red) and theoretical results (blue) of the longitudinal electric field $E_x$ as a function of the central energy of the electron beam. **b** Scaling of the longitudinal electric field $E_x$ versus the electron beam charge.

Poisson equation $\nabla^2 \phi' = -n'_e/\varepsilon_0$, where $\nabla^2 = \partial^2/\partial x'^2 + \partial^2/\partial y'^2$ is the Laplace operator, $\phi'$ and $n'_e$ are potential and electron density in co-moving frame, respectively. Then the electric field in co-moving frame is obtained by $\mathbf{E}' = -\nabla \phi'$. The laboratory results of the longitudinal Coulomb electric field $E_x = E'_x$, whose magnitude is also two orders less than the simulation. Thus, such a strong acceleration field in the simulation stems from the CTR, rather than its own Coulomb field. For transverse electric field $E_y = E_r \cos\varphi$, CTR still accounts for the vast majority.

We plot maximum on-axis value of the longitudinal electric field at different times. The CTR field plus Coulomb field (blue solid circle) from Eq. (1) agree well with the simulation results (red square), as shown in Fig. 3g. We notice that the plasma PIC simulation results agree well with the theory for conductor material. A possible reason is that abundant free electrons exist in both media. In our calculation, we neglect the contribution of positrons to the field, as the initial positron density is lower than the electron density by two orders of magnitudes.

**Discussion.** Further, we explore the acceleration dependence on the drive beam energy by 3D simulations, where the beam profile remains the same as the one in Fig. 3. To compare with the CTR theory, the central energy of the electron beam is tuned from sub-MeV to GeV with mono-energetic spectrum and zero angle divergence. Figure 4a shows the peak value of the acceleration field scaling with the electron energy, obtained at 46.9 μm where the electron beam completely exits the copper target. One sees an uprising in the region of 1–10 MeV and then a saturation to about 100 GV m$^{-1}$ level. The latter comes from the relativistic effect. When $\gamma \gg 1$, the radiation field for an individual charge is $\sim E_x \propto 2\beta e/R$ according to the second term of Eq. (1), i.e., the dependence on the electron energy vanishes. The theoretical calculation from Eq. (1) all matches well with the simulation results in the considered energy region. In Fig. 4b, we show the scaling of the acceleration field as a function of the electron beam charge, while the electron energy is fixed to 500 MeV. We see that the field linearly increases with the charge $Q$, indicating that the transition radiation power is proportional to the square of the number in the electron beam ($W \propto Q^2$), which again clearly confirms that the radiation here is coherent. In terms of the field strength, we find that it is more efficient to increase the beam charge rather than the beam energy, which is why we employed a large focal spot size for the laser pulse to maximize the beam loading in LWFA.

As for the role of the confining magnetic field, we did 2D simulation for stage III when the external magnetic field is 1/100, 1/10, and 1/2 of 30 T. Other parameters are the same with the main 2D simulation. When the external magnetic field reaches beyond 10 T, positrons are well confined as a compact bunch, shown in Supplementary Fig. 2. Essentially, the confining magnetic field strongly depends on the divergence angle of positrons beam, which originally comes from the LWFA electron beam in stage I. The requirements for the external magnetic field will be more tolerant if the electron bunch is more collimated in stage I. In experiments, it is necessary to deplete laser pulse energy (e.g., increase the jet-foil distance or put a laser blocker between stage I and stage II) to avoid the laser–solid interaction[38–41].

In conclusion, we proposed a scheme for GeV-level positron acceleration. The positrons created from the routine BH process in a copper target gain high-energies from the longitudinal electric field induced by the CTR owing to co-propagating electrons exiting into vacuum. Both simulations and theories show that the field approaches tens of GV m$^{-1}$ and sustains for a long distance. Further, increasing the beam charge (rather than the beam energy) is more efficient in enhancing the field strength. Our series of simulations (EPOCH-Geant4-EPOCH) confirm the generation of a quasi-mono-energetic positron beam with central energy at about 500 MeV with an energy spread of 24.3% and a small angular divergence of 2.5 mrad (FWHM). For further acceleration, one could utilize two or more copper targets to provide multistage CTR acceleration. A detailed study will be included in our future work. This all-optical acceleration scheme is available in 100 TW ~ 100 PW laser facilities and offers a promising way to provide high-energy positron sources, which has a wide range of applications in material science, laboratory astrophysics, and potentially, as an injector for future high-energy accelerators.

## Methods

**Stage I: PIC simulation for electron beam acceleration.** This is verified in 2D simulations in a moving window of $L_x \times L_y = 180\lambda \times 360\lambda$ (2000 × 750 cells). Each cell is filled with ten macro particles. A circular-polarized laser pulse (wavelength $\lambda = 0.8$ μm, duration $\tau = 45$ fs) with super-Gaussian with exponent 4 spatial distribution and $\sin^2$ temporal profile propagates along the $x$ direction from the left side. The super-Gaussian profiled high power laser beams, which is a better candidate to laser particle acceleration[42], are being developed in several laboratories[43–46] and 10 PW-level lasers are basically super-Gaussian profile because of the full use of the amplification medium. The dimensionless laser field amplitude is $a_0 = eA/m_e c^2 = 8/\sqrt{2}$ corresponding to an intensity of $1.37 \times 10^{20}$ W cm$^{-2}$, where $A$ is the vector potential, $e$ and $m_e$ are the electron charge and mass respectively. A relatively large focal spot diameter of $r_0 = 60\ \lambda$ is employed to increase the loaded beam charge[47]. This corresponds to peak laser power of 4.95 PW, accessible in facilities such as ELI[48], SULF[46], Apollon[49], Vulcan[50], etc. We use the shock-front injection[51,52] in the gaseous plasma target: the density increases sinusoidally from 0 to $3 \times 10^{18}$ cm$^{-3}$ within 290$\lambda$, then decreases sinusoidally to $2 \times 10^{18}$ cm$^{-3}$ after a short transition length of 10$\lambda$ and maintains afterwards till 13.1 mm.

**Stage II: Monte-Carlo code for positron generation.** Limited by the computational resource, we sample the electrons generated from stage I at $5.34 \times 10^{-4}$ ratio (~$10^7$ simulation particles) by reserving the energy and angular spectrum and input them in the Geant4 simulations. The copper target with thickness of 1 mm is set 0.5 mm away from the gas target. The detection device is at the rear side of the copper target.

**Stage III: PIC simulation for positron acceleration.** The acceleration process is verified by 2D PIC simulation using a moving window of 40 μm × 400 μm divided into 400 × 1200 cells with 5 macro particles per cell for both electrons and positrons. The hybrid beam taken from stage II is initialized in the copper target. The plasma density is $n_{Cu} = 8.49 \times 10^{22}$ cm$^{-3}$ and covers the region 0 < $x$ < 20 μm. The initial electron beam containing 3 nC beam charge is approximated by a Gaussian density profile $n_e = n_{e0}\exp(-x^2/\delta_{ex}^2 - y^2/\delta_{ey}^2)$, where $n_{e0} = 1.46 \times 10^{18}$ cm$^{-3}$, $\delta_{ex} = 3.18$ μm ($\delta_{FWHM} = 5.3$ μm), and $\delta_{ey} = 26.9$ μm ($\delta_{FWHM} = 44.8$ μm), respectively, according to the numbers in stage II. The 12.1 pC positron beam is profiled with $\delta_{px} = 3.18$ μm ($\delta_{FWHM} = 5.3$ μm) and $\delta_{py} = 25.5$ μm ($\delta_{FWHM} = 42.5$ μm), but a lower density of $n_{p0} = 6.5 \times 10^{15}$ cm$^{-3}$. Electron energy distribution is taken from stage II. Here we have considered the transportation effect in the copper foil for electron spectrum initially shown by the black solid line in Fig. 1b. Positrons have an initial Maxwellian energy distribution at temperature $T_p = 24.1$ MeV. We apply an external longitudinal magnetic field of $B_x = 30$ T to guide the electron/positron beam propagating in vacuum for long distances. Such fields are now available using superconducting material[53].

## Data availability

The data that support the findings of this study are available from the corresponding authors on reasonable request.

### Acknowledgements
This work was supported by Ministry of Science and Technology of the People's Republic of China (2018YFA0404803 and 2016YFA0401102), Strategic Priority Research Program of the Chinese Academy of Sciences (XDB16), the Scientific Equipment Research Project of Chinese Academy of Sciences (Project number 1701521X00) and the National Natural Science Foundation of China (Project numbers 11935008, 11775287, 11875307, and 11705261).

### Author contributions
B.F.S. designed the project. Z.L.X. performed the simulations, calculations, and drafted the manuscript. L.Q.Y. and L.L.J. contributed to the key physical mechanism. L.G.Z. and S.L. assisted with the simulations. L.L.J., J.C.X., and B.F.S. contributed to the revisions. T.J.X. and Z.Z.X. gave useful advices. All authors discussed the results and commented on the manuscript.

### Competing interests
The authors declare no competing interests.

### Additional information
**Supplementary information** is available for this paper at https://doi.org/10.1038/s42005-020-00471-6.

**Correspondence** and requests for materials should be addressed to B.S., J.X. or L.J.

**Reprints and permission information** is available at http://www.nature.com/reprints

**Publisher's note** Springer Nature remains neutral with regard to jurisdictional claims in published maps and institutional affiliations.